\documentclass{aastex}
\usepackage{spr-astr-addons}
\RequirePackage{color}

\begin{document}

\title{Variation of Bar Strength with Central Velocity Dispersion in Spiral Galaxies}

\shorttitle{Bar Strength and Central Velocity Dispersion}
\shortauthors{Das et al.}

\author{M.Das\altaffilmark{1}} \affil{Raman Research Institute, Bangalore, 560080, India.\\
E-mails~:~mousumi{@}rri.res.in}
\and \author{E.Laurikainen\altaffilmark{2}} \and \author{H.Salo\altaffilmark{2}}
\affil{Division of Astronomy, Department of Physical Sciences, University of Oulu, FIN-90014, Finland}
\and \author{R.Buta\altaffilmark{3}}
\affil{Department of Physics and Astronomy, University of Alabama, Box 870324, Tuscaloosa, AL 35487, U.S.A.}


\begin{abstract}
We investigate the variation of bar strength with central velocity
dispersion in a sample of barred spiral galaxies. The bar strength is
characterized by $Q_g$, the maximal tangential perturbation
associated with the bar, normalized by the mean axisymmetric
force. It is derived from the galaxy potentials
which are obtained using near-infrared images of the galaxies.
However, $Q_g$ is sensitive to bulge mass. Hence we also estimated
bar strengths from the relative Fourier intensity amplitude ($A_{2}$) of bars in near-infrared images.
The central velocity dispersions were obtained from integral field spectroscopy observations
of the velocity fields in the centers of these galaxies; it was
normalized by the rotation curve amplitude obtained from HI line width for each galaxy. We found a
correlation between bar strengths (both $Q_g$ and $A_{2}$) and the normalized
central velocity dispersions in our sample. This suggests
that bars weaken as their central components become kinematically hotter. This may have important
implications for the secular evolution of barred galaxies.
\end{abstract}



\section{INTRODUCTION}

In past decades there has been a large amount of discussion regarding the secular evolution of bars
in spiral galaxies. Numerical simulations show that gas inflow, star formation and the subsequent
buildup of a central mass concentration (CMC) can considerably alter the shape of a bar and even result
in the dissolution of the bar itself (Norman, Sellwood \& Hasan 1996;
Shen \& Sellwood 2004; Kormendy \& Kennicutt 2004; Hozumi \& Hernquist 2005; Athanassoula, Lambert \&
Dehnen 2005; Bournaud, Combes \& Semelin 2005).
However, the fraction of bars remains
fairly high out to a redshift of z$\sim$0.7 to 1 (Sheth et al. 2003; Elmegreen, Elmegreen \& Hirst 2004;
Jogee et al. 2004; Sheth et al. 2007) which suggests that either bars are long-lived or can reform in galaxy disks.
Simulations suggest that bars can dissolve but
form again once the disk is cooled through gas accretion
(Bournaud \& Combes 2002) or through tidal interactions (Berentzen et al. 2004).

Although there has been a considerable amount of numerical work on bar formation and dissolution,
there is not much observational evidence. Molecular gas
has a higher central concentration in barred galaxies compared to non-barred galaxies which suggests
that bars are important for building up a CMC
(Sakamoto et al. 1999; Sheth et al. 2005). The bar length has been found to correlate with bar intensity contrast
and central luminosity densities (Elmegreen et al. 2007). A more direct indicator of how
bars change with mass concentration is the
correlation of CMC and bar ellipticity (Das et al. 2003) which shows that bar ellipticity declines with
increasing dynamical CMC. An enhanced CMC also results in the
deepening of the central potential and gives rise to a dynamically hotter galaxy center. This
will lead to a higher observed central stellar velocity dispersion in a galaxy ($\sigma_{v}$).
This effect has been seen in
numerical simulations where a dynamically hotter center results when a bar dissolves
(Friedli \& Benz 1993; Hasan, Pfenniger \& Norman 1993) or weakens (Athanassoula, Lambert \& Dehnen 2005).
In this paper we investigate whether there is any observational evidence for this heating effect by
examining the correlation between the central velocity dispersion
in bars and the bar strength.

\section{Central Velocity Dispersion}

The $\sigma_{v}$ values used in this paper are from published integral-field spectroscopy (IFS)
observations of the nuclear regions of galaxies. We have thirty one galaxies in our sample (Table~1).
The majority of $\sigma_{v}$ values are from the SAURON survey of nearby
galaxies (de Zeeuw et al. 2002; Ganda et al. 2006; Falcon-Barroso et al. 2006; Peletier et al. 2007); a smaller
number are from
the INTEGRAL and SPIRAL instruments on the William Herschel and Anglo-Autralian Telescopes
respectively (Batcheldor et al. 2005). The $\sigma_{v}$ values for the remaining four galaxies
are from observations using the GMOS instrument at the Gemini North Telescope (Barbosa et al. 2002).
The velocity dispersion values that we have used are derived from two dimensional stellar velocity
fields and not gas kinematics.

One drawback of using these $\sigma_{v}$ values is that the aperture size
is not uniform across the sample. Hence we have tried to bring the apertures to a common system
using the aperture correction formula of Jorgensen, Franx \& Kjaergaard (1995). The velocity dispersions
were corrected to a radius $R_{e}/8$, where $R_{e}$ is the effective bulge radius of a galaxy.
Thus $\sigma_{v}$ was assumed equal to the average velocity dispersion within a radius $R_{e}/8$ of the
bulge of a galaxy.
For five galaxies it was not possible to determine $R_{e}$ either because the
bulge was not distinguishable or the image quality was poor. Also, since the galaxies span a wide variety
of size and mass, we normalised the new aperture corrected velocity dispersions ($\sigma_{e}$)
with the HI gas rotation velocity $v_{g}$ for each
galaxy. This also takes out some of the luminosity/size effects for the galaxies. The $v_{g}$ values
were derived from the maximum HI rotation velocities ($v_{H}$) using the galaxy axes ratios ($q$)
to correct for inclination (i.e. $v_{g}=v_{H}/\sqrt{1-q^{2}}$~); $v_{H}$ was obtained from the
Hyperleda database.
The ratio $\sigma_{e}/v_{g}$ is thus an indicator
of how kinematically hot the center is for each galaxy relative to its disk rotation speed.

\section{Bar Strength Derived from Near-IR Images}

There are several ways to quantify the strength of a bar in a
galaxy. In this paper we have derived the bar strength in two ways;
bar strength is assumed to be the maximum relative bar torque ($Q_g$) derived
from the gravitational potential of a galaxy. This is probably the
most robust estimate of bar strength but it is sensitive to the bulge
mass or luminosity. The second way of estimating bar strength is to
use the maximum of the relative intensity amplitude of the bar in the near-IR image
($A_2$). Both methods require that the images be deprojected before analysis.
This assumes the disks are thin, but the presence of a less flattened
bulge component could lead to artificial stretching of bulge isophotes.
To minimize this effect, we decomposed the bulge from the disk,
subtracted it from the image before deprojection, and added it back
(assuming that bulge light is spherically symmetric)
after deprojecting the disk/bar light. The bulge components were separated
from the disks using a two-dimensional multicomponent
decomposition code which uses a
Sersic model for the bulge, an exponential function for the disk,
and either Ferrers' or Sersic's functions for the bar (Laurikainen, Salo, and Buta 2005).
Effects of seeing were taken into account, using the values of the full width at half maximum stored
in image headers or provided in articles. The effective radius of the bulge ($R_e$) was estimated by
integrating the flux of the fitted bulge model. The images were obtained mainly from the 2MASS survey and
some from previous studies. The fiters used were either K or H band. The bar parameters were derived 
as follows.

\noindent
{\bf (i)~$Q_g$ :}~The gravitational potentials ($\Phi$) were inferred from near-IR light
distributions assuming that the light traces the mass. Bar induced tangential forces were calculated using
a Polar method, as described in Laurikainen \& Salo (2002) and Laurikainen, Salo \& Buta (2004).
In particular, the
calculation applies an azimuthal Fourier decomposition of intensity,
including the even components up to m=20, which are then converted to
the corresponding potential components (Salo et al. 1999).
Two dimensional maps of the radial force
($F_R$) and tangential force ($F_T$) were calculated. The radial profile
of the maximum tangential force at each distance is given by,

\begin{equation}
Q_T(r) = { ~~~~~|F_{T}(r,\phi)|_{max} \over <|F_{R}(r,\phi)|>}
\end{equation}

\noindent
where  $<|F_{R}(r,\phi)|>$ denotes the azimuthally averaged axisymmetric
force at each radius. The maximum in the $Q_T$ profile at the region of the bar
then gives a single
measure of bar strength, $Q_g$. The main assumptions are that the
mass-to-luminosity ratio is constant in the bar region, and that the
vertical light distribution can be approximated by an exponential.
The scale height, $h_z$, was estimated from an empirical relation
between $h_r$/$h_z$ and the de Vaucouleurs type index T (de Grijs 1998), where
$h_r$ is the radial scale length of the disk.
The images were taken from the literature and are generally not very deep.
Therefore, instead of estimating $h_r$ from the new decompositions, we used mainly the
$h_r$ values from Baggett et al. (1998); if it was not available, we
used other sources in the literature where deep optical images had been used to derive
$h_r$. For two
galaxies $h_r$ was taken to be the mean value for our sample
of 31 galaxies.

\noindent
{\bf (ii)~$A_2$ :}~The same Fourier method gives us also the m=2 amplitudes of
bar intensity contrast in the bar region.
For some of the galaxies, Table 1 gives $Q_g$ and $A_2$ but not the effective bulge radius.
This is because application of the 2D decomposition method requires deeper images than
the methods used to calculate $Q_g$ and $A_2$.

\vspace{-2mm}
\section{Estimating the Correlation }

Figure~1 shows bar strength $Q_g$ plotted against the normalized
velocity dispersion $\sigma_{e}/v_g$ for 26 galaxies. Although there are 31 galaxies in the sample,
we could use only 26 because the effective radius $R_e$ could not be calculated for a few cases.
The errors have been calculated using the standard error propagation equation and include
the uncertainties in the observed quantities. The majority of galaxies seem to follow a trend of
decreasing bar strength with increasing central velocity dispersion.
We quantified the correlation in two ways.
The linear correlation coefficient for this sample of 26 galaxies
is $r=-0.50$ and the probabilty that
they are from a random sample $P_r$ is $P_r~<~1$\%.
This method does not include the errors on both axes. A more accurate
estimate would be a weighted correlation coefficient, but this is difficult to obtain in practice
(Feigelson \& Babu 1992). Instead we used a simple Monte Carlo simulation that
randomly samples the errors on both axes
and determines a mean weighted correlation coefficient $<r>$. We used 50,000 linear fits
and obtained a value of $<r>=-0.46$.
The second method that we used to quantify the correlation was the Kendall-Tau coefficient which assigns
relative ranks to the different values. This is perhaps a more robust way of examining the correlation
especially when the sample size is relatively small as in our case.
The Kendall-Tau coefficient for the 26 galaxies in Figure~1 is
$<r_{KT}>=-0.35$ and the probability that they are from a random sample is $P_{KT}\sim1.3$\%.

Figure~2 shows the plot of $A_{2}$ against the normalized velocity dispersion $\sigma_{e}$
for 25 galaxies. Here again we calculated the linear correlation coefficient which is $r=-0.54$ and
$P_r~<~0.5$\%. When the errors are sampled using a
Monte Carlo simulation we obtain a value of $<r>=-0.51$. The Kendall-Tau coefficient for the 25
galaxies is $<r_{KT}>=-0.33$ and $P_{KT}~<~2$\%.
Thus both Figures~1 and 2 suggest that there is a correlation between the bar strength in
galaxies and their central velocity dispersions.

\section{DISCUSSION}

The main result of this paper are shown in Figures 1 \& 2. 
There are 26 galaxies in the plots of which about half are intermediate type
spirals and the remaining a mixture of early and late type spirals. Since the number of galaxies in 
each Hubble type is not very large, we cannot investigate trends within the different Hubble types. 
But from Figures 1 \& 2, it appears that early type spirals have relatively lower central dispersions;
this may be because they have larger bulges where rotational velocity is comparatively higher and
the central velocity dispersion lower compared to the later Hubble types.

Since the correlation is significant but not very strong we examined the two galaxies that define 
the higher and lower limits of $Q_g$ and $A_2$. We looked at them closely to see if they are odd in 
some way, and not characteristic of the rest of the sample. (i)~NGC~3162 ($\sigma_{e}/v_g$=1.13)~:~This
is an intermediate type spiral galaxy with a weak bar and prominent bulge. There may also be a ring in 
the center. Though the spiral arms are somewhat asymmetric, the nucleus is fairly undisturbed. 
(ii)~NGC~4314 ($\sigma_{e}/v_g$=0.18)~:~This is a bright, early type galaxy with a strong bar and large
bulge with a LINER type nucleus. There may be significant rotation in the nuclear region which may lower
the $\sigma_{e}$. Also, the mass may be more widely distributed over the bucle and hence not as centrally 
concentrated as in NGC~3162. There appears to be a ring of star formation in the nucleus as well 
(Gonzalez Delgado et al. 1997). Both galaxies are thus fairly normal and not unusually different from
the rest of the sample galaxies.  

Figures~1 \& 2 suggest that galaxies with dynamically hotter nuclei have weaker bars. It is
now well established that a galaxy's central black hole mass and bulge velocity dispersion are correlated
(Ferrarese \& Merritt 2000; Gebhardt et al. 2000). Later results show that the nuclear mass
also correlates with the overall mass of a galaxy (Ferrarese et al. 2006; Hopkins et al. 2007).
These results all indicate that the nuclear mass in a galaxy is intimately connected to the dynamics of
its disk and halo. If so, then it is not suprising that $\sigma_{e}$ in our sample of barred galaxies is
correlated with $Q_g$ or $A_{2}$. It suggests that the growth of a central mass and evolution of the bar for
spiral galaxies may be closely linked.

Our results have important implications for the secular evolution of barred galaxies. Simulations
suggest that there may be several factors responsible for the dissolution of a bar. One is the CMC growth
(Hasan \& Norman 1990; Friedli \& Pfenniger 1991) which weakens the bar-supporting $x_1$ orbits and increases
the fraction of chaotic orbits in the galaxy center. Second
is the inflow of gas towards the galaxy center
which results in the transfer of angular momentum to the bar wave which then weakens the bar
itself (Bournaud, Combes \& Semelin 2005).
The buckling instability is an important bar thickening mechanism that results
in a boxy/peanut bulge which can temporarily weaken a bar (Raha et al. 1991; Berentzen et al. 1998;
Athanassoula \& Misiriotis 2002; Martinez-Valpuesta, Shlosman \& Heller 2006; Debattista et al. 2006). All these
effects result in a more massive and dynamically hotter central component and a weaker bar.
The correlations that we see in Figures 1 and 2 may be observational
indications of this ongoing evolution. In particular, while the
apparent drop of $Q_g$ with central velocity dispersion might be an artifact caused by
the bulge dilution effect (see Laurikainen, Salo \& Buta 2004; a bias
could follow since nuclear velocity dispersion and bulge mass are strongly correlated), 
the similar correlation between the bar intensity
contrast $A_2$ suggest that the effect is real. The found correlation
between $A_2$ and $\sigma_e/v_g$ is also consistent with Das et al. (2003)
who found that the bar ellipticity (closely related to $A_2$) drops with
central mass concentration.

The evolution of bars by secular processes in galaxies is an issue
which is expected to gain more attention in the near future.
Recent observational evidence shows that the
fraction of strong bars in bright galaxies increases from
under 10\% at redshift z=0.84 to about 30\% in the local universe (Sheth et 2007).
Also, it has been shown by (Laurikainen et al. 2007) that
among the early-type barred galaxies the bulge-to-total flux
ratios are on average smaller than in the non-barred galaxies. These
results, together with ours, may indicate that bars
evolve with their parent galaxies.

\section{ACKNOWLEDGMENTS}

This research has
made use of the NASA/IPAC Infrared Science Archive (NED), which is operated
by the JPL, California Institute of Technology,
under contract with NASA. We also acknowledge the usage of the HyperLeda database
(http://leda.univ-lyon1.fr). R.B. is supported by NSF grant AST 05-07140.
E.L and H.S acknowledge the support by the Academy of Finland.

\nocite{*}
\bibliographystyle{spr-mp-nameyear-cnd}
\bibliography{biblio-u1}

\clearpage

\begin{table*}
\tabletypesize{\scriptsize}
\caption{Galaxy Data}
\begin{tabular}{@{}crrrrrrrrrrr@{}}
\tableline
Galaxy & Class & Type & Galaxy & Bar
& Bar & Nucl. & App. & Bulge
& HI Peak & Normal. & $\sigma_{v}$\\
Name &  &  & Axes & Tor- 
& Amp. & Vel. & Size & Rad-
& Vel. & Disp. & ref.\\
& & & Ratio & que & & Disp. & ($\arcsec$) & ius($\arcsec$) & (km/s) & (km/s) & \\
\tableline
NGC~0289 & SB(r$\underline{\rm s}$)bc & inter. & 0.79 & 0.21 & 0.39 & 114 & 2.7 & 2.35 & 125.6 & 0.59 & (a) \\
NGC~0613 & SB(rs)bc     & inter. & 0.772  & 0.40 & 0.75 & 99 & 2.7 & 4.17 & 168.6 & 0.39 & (a) \\
NGC~0864 & S$\underline{\rm A}$B(rs)c & late & 0.842  & 0.36 & 0.44 & 65 & 2.4 & 1.83 & 98.2 & 0.38 & (b) \\
NGC~1255 & SAB(rs)bc    & inter. & 0.631  & 0.14 & 0.22 & 69 & 2.7 & 0.47 & 113.8 & 0.53 & (a) \\
NGC~1300 & SB(s)b  & inter. & 0.760 & 0.54 & 0.74 & 90 & 2.7 & 3.41 & 126.7 & 0.48 & (a) \\
NGC~1832 & SB(r)bc & inter. & 0.702 & 0.20 & 0.41 & 102 & 2.7 & 1.83 & 123.4 & 0.63 & (a) \\
NGC~2273 & (RR)SAB(rs)a & early & 0.596 & 0.20 & 0.57 & 104 & 1.0 & 2.64  & 161.8 & 0.53 & (d) \\
NGC~2805 & SAB(rs)d & late & 0.758 & 0.26 & 0.57 & 46 & 2.4 & ....  &  41.6 & .... & (b) \\
NGC~2903 & SAB(rs)bc & inter. & 0.471 & 0.30 & 0.50 & 95 & 2.7 & ....  & 173.8 & .... & (a) \\
NGC~2964 & S$\underline{\rm A}$B(r$\underline{\rm s}$)b & inter. & 0.566  & 0.31 & 0.42 & 101 & 2.4 & 1.41 & 134.1 & 0.67 & (b) \\
NGC~3162 & SAB(rs)bc & inter. & 1.000  & 0.12 & 0.21 & 85  & 2.7 & 1.39 &  81.4 & 1.13 & (a) \\
NGC~3227 & SAB(s)a pec & early & 0.661  & 0.16 & 0.44 & 114 & 1.0 & 1.81 & 120.9 & 0.73 & (d) \\
NGC~3346 & SB(rs)cd   & late & 0.860  & 0.41 & 0.30 & 48  & 2.4 & ....  &  69.3 & .... & (b) \\
NGC~3953 & SB(r)bc & inter. & 0.466  & 0.15 & 0.42 & 146  & 2.7 & ....  & 190.8 & .... & (a) \\
NGC~4051 & SAB(rs)c & late & 0.846  & 0.28 & 0.66 & 85 & 1.0 & 3.22 & 107.3 & 0.43 & (d) \\
NGC~4088 & SAB(s)c pec & late & 0.389  & 0.40 & 0.43 & 87   & 2.7 & 2.53 & 158.4 & 0.54 & (a) \\
NGC~4102 & SAB(s)b? & inter. & 0.575  & 0.11 & 0.49 & 150 & 2.4 & 1.86 & 134.4 & 0.98 & (b) \\
NGC~4245 & SB(r)0/a & early & 0.823  & 0.19 & 0.54 & 36 &  2.4 & 4.24 & 94.2 & 0.22 & (c) \\
NGC~4258 & (R$^{\prime}$)SAB(r$\underline{\rm s}$)b & inter. & 0.389 & 0.30 & 0.57 & 120  & 2.7 & 9.94 & 197.8 & 0.56 & (a) \\
NGC~4274 & (R$^{\prime}$)SB(r)ab & inter. & 0.391 & 0.34 & 0.62 &  99 & 2.4 & ....  & 219.5 & .... & (c) \\
NGC~4293 & (R)SB(s)0/a  & early &  0.463  & 0.36 & 0.67 & 42  & 2.4 & 4.65 & 109.5 & 0.35 & (c) \\
NGC~4303 & SAB(rs)bc & inter. & 0.861  & 0.26 & 0.44 & 108  & 2.7 & 2.96 & 66.3 & 0.87 & (a) \\
NGC~4314 & (R$_1^{\prime}$)SB(r$^{\prime}$l)a & early & 0.959 & 0.44 & 0.90 & 43 & 2.4 & 5.46 & 70.7 & 0.18 & (c) \\
NGC~4321 &  SAB(s)bc    & inter. & 0.866  & 0.18 & 0.34 & 101  & 2.7 & 8.62 & 112.5 & 0.45 & (a) \\
NGC~4487 & SAB(rs)cd    & late & 0.659  & 0.18 & 0.18 & 51 & 2.4 & 29.14 & 86.7 & 0.42 & (b) \\
NGC~4593 & (R$^{\prime}$)SB(rs)ab & inter. & 0.742 & 0.31 & 0.76 & 105 & 1.0 & 4.32 & 161.3 & 0.44 & (d) \\
NGC~4596 & SB(rs)0/a    & early & 0.716  & 0.28 & 0.67 & 59  & 2.4 & 2.78 & 92.5 & 0.47 & (c) \\
NGC~5005 & SAB(s)b      & inter. & 0.444  & 0.15 & 0.36 & 203 & 2.7 & 3.03 & 243.8 & 0.79 & (a) \\
NGC~5448 & (R$_1^{\prime}$)S$\underline{\rm A}$B($\underline{\rm r}$s)ab & inter. & 0.454  & 0.15 & 0.65 & 65  & 2.4 & 2.83 & 188.4 & 0.32 & (c) \\
NGC~5585 & (R$^{\prime}$)SAB(s)d & late & 0.646  & 0.24 & .... & 42 & 2.4 & 0.17 & 63.4 & 0.59 & (b) \\
NGC~5678 & SAB(rs)b & inter. & 0.489  & 0.18 & 0.33 & 103 & 2.4 & 2.43 & 175.2 & 0.54 & (b) \\
\tableline
\end{tabular}
\tablenotetext{a}{(i)~deVA= de Vaucouleurs Atlas of Galaxies (Buta et al. 2007)}
\tablenotetext{b}{(ii)~RC3= Third Reference Catalogue of Bright Galaxies (de Vaucouleurs, et al. 1991)}
\tablenotetext{c}{(iii)~a=Batcheldor et al. (2005); b=Ganda et al. (2006); c=Peletier et al. (2007)
d=Barbosa et al. (2006)}
\end{table*}

\begin{figure}[t]
\includegraphics[width=80mm,height=80mm,angle=-90]{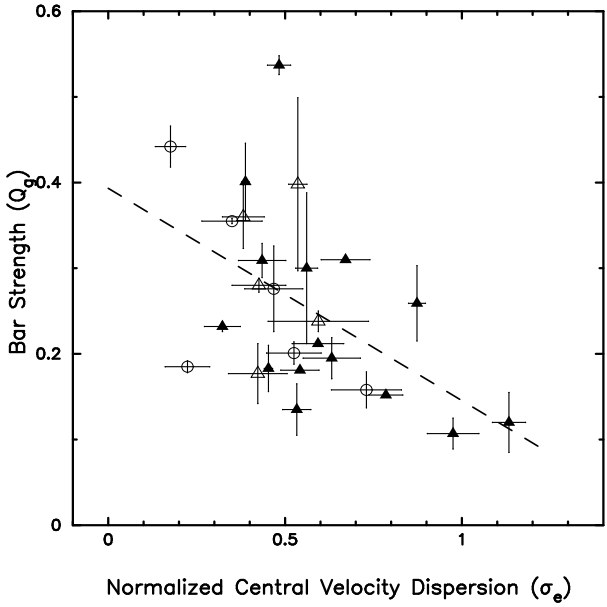}
\caption{The plot shows the bar strength $Q_{g}$ plotted against the normalized central velocity
dispersion ($\sigma_{e}$/$v_{g}$) for  26 galaxies from Table~1.
The three galaxy types are plotted with different symbols; early-type spirals are S0/a to Sa (open circles),
intermediate-type spirals are Sab to Sbc (filled triangles) and the late-type spirals are Sc to Sd
(open triangles).  The errors are marked on each axis.
The dashed line is the best fit line and has the form $y=-0.25x + 0.39$.
Although there is some scatter, the overall trend suggests that bars weaken as their nuclei become
dynamically hotter. }
\end{figure}

\vspace{30mm}

\begin{figure}[t]
\includegraphics[width=80mm,height=80mm,angle=-90]{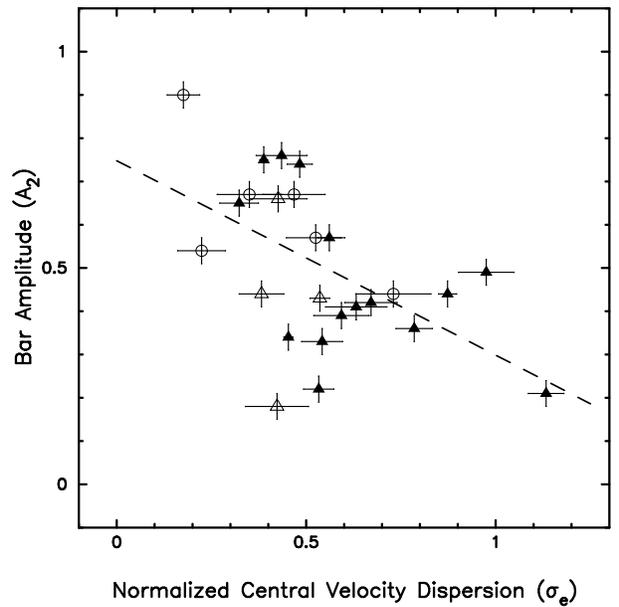}
\caption{The plot shows bar amplitude $A_2$ measured from the near-IR images plotted
against the normalized central velocity dispersion ($\sigma_{e}$/$v_{g}$) for a sample of 25 galaxies.
As in Figure~1 the open circles
represent early type spirals, the filled triangles represent intermediate type spirals
and the open triangles represent the late type spirals.
The dashed line is the best fit line for the sample and has a form y=-0.45x + 0.75. }
\end{figure}

\end{document}